\documentclass{whitepaper}

\usepackage[dvipsnames]{xcolor}
\definecolor{color1}{HTML}{1f77b4}
\definecolor{color2}{HTML}{ff7f0e}
\definecolor{color3}{HTML}{2ca02c}
\definecolor{color4}{HTML}{d62728}

\definecolor{kneered}{HTML}{EE3333}
\definecolor{thresh90}{HTML}{EE7733}
\definecolor{thresh95}{HTML}{009988}
\definecolor{thresh99}{HTML}{AA3377}
\definecolor{almaorange}{HTML}{CA6641}

\usepackage{acro}

\usepackage{cleveref}
\usepackage{xspace}
\acsetup{
  single    = false,
  list/sort  = true,
  cite/group = true,
  cite/group/cmd = \citep,
  cite/group/pre = {; \xspace},
  patch/longtable=false,
}

\newcommand{\amigo}{\textsc{Amigo}\xspace}

\newcommand\stpsf{\texttt{stpsf}\xspace}

\newcommand\nacho{\textsc{Nacho}\xspace}
\newcommand{\dorito}{\texttt{dorito}\xspace}
\newcommand{\frito}{\texttt{frito}\xspace}
\newcommand{\jax}{\textsc{Jax}\xspace}

\DeclareAcronym{amigo}{
  short = \amigo,
  long = Aperture Masking Interferometry Generative Observations,
  cite = {amigo}
}
\DeclareAcronym{psf}{
  short = PSF,
  long = Point Spread Function
}
\DeclareAcronym{otf}{
  short = OTF,
  long = Optical Transfer Function
}
\DeclareAcronym{ifu}{
  short = IFU,
  long = Integral Field Unit
}
\DeclareAcronym{fft}{
  short = FFT,
  long = Fast Fourier Transform
}
\DeclareAcronym{rom}{
  short = ROM,
  long = Reduced Order Modelling,
  cite = {Benner2015}
}
\DeclareAcronym{autodiff}{
  short = autodiff,
  long = Automatic Differentiation,
  cite={autodiff}
}
\DeclareAcronym{jwst}{
  short = JWST,
  long = \textit{James Webb Space Telescope}
}
\DeclareAcronym{niriss}{
  short = NIRISS,
  long = Near Infrared Imager and Slitless Spectrograph
}
\DeclareAcronym{nir}{
  short = NIR,
  long = Near Infrared
}
\DeclareAcronym{ami}{
  short = AMI,
  long = Aperture Masking Interferometer,
  cite = {Sivaramakrishnan_2023}
}
\DeclareAcronym{bfe}{
  short = BFE,
  long = Brighter-Fatter Effect,
  cite = {bfe_ccd,bfe_ccd2,bfe_miri,bfe_fgs,bfe_pipeline}
}
\DeclareAcronym{edm}{
  short = EDM,
  long = Effective Detector Model
}
\DeclareAcronym{nn}{
  short = NN,
  long = Neural Network
}
\DeclareAcronym{cnn}{
  short = CNN,
  long = Convolutional Neural Network
}
\DeclareAcronym{ml}{
  short = ML,
  long = Machine Learning
}
\DeclareAcronym{fov}{
  short = FOV,
  long = field of view
}

\DeclareAcronym{h2rg}{
  short = H2RG,
  long = Teledyne HAWAII-2RG
}
\DeclareAcronym{adc}{
  short = ADC,
  long = Analogue to Digital Converter
}
\DeclareAcronym{ipc}{
  short = IPC,
  long = Inter-Pixel Capacitance
}
\DeclareAcronym{disco}{
  short = DISCO,
  long = Delay-Invariant Subspace of Calibrated Observables
}
\DeclareAcronym{nrm}{
  short = NRM,
  long = Non-Redundant Mask
}
\DeclareAcronym{hst}{
  short = HST,
  long = \textit{Hubble Space Telescope}
}
\DeclareAcronym{mcmc}{
  short = MCMC,
  long = Monte Carlo Markov Chain,
  cite={Metropolis1953}
}
\DeclareAcronym{iwa}{
  short = IWA,
  long = Inner Working Angle
}
\DeclareAcronym{opd}{
  short = OPD,
  long = Optical Path Difference
}
\DeclareAcronym{go}{
  short = GO,
  long = General Observer
}
\DeclareAcronym{sgd}{
  short = SGD,
  long = stochastic gradient descent,
  cite= {Ruder2016}
}
\DeclareAcronym{hmc}{
  short = HMC,
  long = Hamiltonian Monte Carlo,
  cite= {Betancourt2017}
}
\DeclareAcronym{mast}{
  short = MAST,
  long = Mikulski Archive for Space Telescopes
}

\DeclareAcronym{gpu}{
  short = GPU,
  long = Graphics Processing Unit
}

\DeclareAcronym{nicmos}{
  short = NICMOS,
  long = Near-Infrared Camera and Multi-Object Spectrometer
}

\DeclareAcronym{adi}{
  short = ADI,
  long = Angular Differential Imaging,
}

\DeclareAcronym{rdi}{
  short = RDI,
  long = Reference Differential Imaging
}

\DeclareAcronym{kpi}{
  short = KPI,
  long = Kernel Phase Interferometry,
  cite = {kernel_phase}
}

\DeclareAcronym{ota}{
  short = OTA,
  long = Optical Telescope Assembly,
}

\DeclareAcronym{nacho}{
  short = \nacho,
  long = NICMOS Aperture Characterisation to High Order
}
\DeclareAcronym{mft}{
  short = MFT,
  long = Matrix Fourier Transform,
  cite= {soummer_mft}
}

\DeclareAcronym{lct}{
  short = LCT,
  long = Linear Canonical Transform,
}

\DeclareAcronym{pam}{
  short = PAM,
  long = Pupil Alignment Mechanism,
}

\DeclareAcronym{nuts}{
    short = NUTS,
    long = No U-Turns Sampler,
    cite = {nuts}
}

\DeclareAcronym{dorito}{
    short = \dorito,
    long = Differentiable Optics for Reconstructing Images from Telescope Observations,
    cite = {dorito}
}

\DeclareAcronym{frito}{
    short = \frito,
    long = Feature-space Reconstruction for Images from Telescope Observations
}

\DeclareAcronym{ao}{
    short = AO,
    long = adaptive optics,
}

\DeclareAcronym{epsf}{
  short = ePSF,
  long = effective Point Spread Function,
  cite = {Anderson_king_ePSF}
}

\DeclareAcronym{chirp}{
    long = Continuous High-resolution Image Reconstruction using Patch priors,
    short = CHIRP,
    cite = {Bouman_CHIRP}
}

\DeclareAcronym{vlbi}{
    long = Very Long Base Interferometer,
    short = VLBI
}

\DeclareAcronym{jvm}{
    long =  Jorsater \& van Moorsel,
    short = JvM,
    cite = {JvM}
}

\DeclareAcronym{iris}{
    long = Imaging for Radio Interferometry with Score-based models,
    short = IRIS,
    cite = {Dia2025}
}

\DeclareAcronym{agn}{
    long = Active Galactic Nuclei,
    short = AGN
}

\DeclareAcronym{cmos}{
    long = Complementary metal-oxide-semiconductor,
    short = CMOS
}

\DeclareAcronym{ccd}{
    long = Charge-Coupled Device,
    short = CCD
}

\DeclareAcronym{mem}{
    long = Maximum Entropy,
    short = MaxEnt
}

\DeclareAcronym{tv}{
    short = TV,
    long = Total Variation,
}

\DeclareAcronym{dip}{
    long = Deep Image Prior,
    short = DIP,
    cite = {Ulyanov_2020, Ren_2020}
}

\DeclareAcronym{rn}{
    long = Residual Neural Network,
    short = ResNet
}

\DeclareAcronym{svd}{
    long = Singular Value Decomposition,
    short = SVD
}

\DeclareAcronym{pca}{
    long = Principle COmponent Analysis,
    short = PCA
}

\DeclareAcronym{bfgs}{
    short = BFGS,
    long = Broyden\textendash\allowbreak Fletcher\textendash\allowbreak Goldfarb\textendash\allowbreak Shanno,
    cite = {bfgs1, bfgs2, bfgs3, bfgs4}
}

\DeclareAcronym{map}{
    long = maximum \textit{a posteriori},
    short = MAP
}

\DeclareAcronym{snr}{
    long = Signal-to-noise ratio,
    short = SNR
}

\coverheader{%
  Multi-Cycle Treasury Program Concepts with JWST: A White Paper\\
  \today } 

\coverauthors{%
  \au[0000-0003-2595-9114]{Benjamin Pope}{$*$,1},
  \au[0000-0002-2106-0403]{Yinuo Han}{$*$,2},
  \au[0000-0003-2253-2270]{Sean Andrews}{3},
  \au[0000-0002-3780-4112]{Kateryna Andrych}{1},
  \au[0000-0001-5032-1396]{Phil Armitage}{4},
  \au[0000-0001-6396-8439]{William Balmer}{5},
  \au[0000-0002-7695-7605]{Myriam Benisty}{6},
  \au[0000-0002-3968-3780]{Faustine Cantalloube}{7},
  \au[0009-0003-5950-4828]{Max Charles}{8},
  \au[0000-0003-2076-8001]{Ilse Cleeves}{9},
  \au[0000-0002-1483-8811]{Ian Czekala}{10},
  \au[0000-0002-1015-9029]{Louis Desdoigts}{11},
  \au[0000-0001-9290-7846]{Ruobing Dong}{12},
  \au[0000-0002-2683-2396]{Rodrigo Ferrer-Chavez}{5},
  \au[0000-0002-7821-0695]{Katherine Follette}{13},
  \au[0000-0002-4438-1971]{Christian Ginski}{14},
  \au[0000-0001-7138-0563]{Jonah Goldfine}{1},
  \au[0009-0008-8188-8668]{Hayden Greer}{8},
  \au[0000-0002-8138-0425]{Cassandra Hall}{15},
  \au[0000-0001-8074-2562]{Sasha Hinkley}{16},
  \au[0000-0003-2948-5614]{Haochang Jiang}{6},
  \au[0000-0001-8299-3402]{Devika Kamath}{1},
  \au[0000-0003-2769-0438]{Jens Kammerer}{17},
  \au[0000-0002-7064-8270]{Matthew Kenworthy}{11},
  \au[0009-0000-9575-6190]{Milo Langker}{8},
  \au[0000-0002-6964-8732]{Kellen Lawson}{18},
  \au[0000-0003-2458-9756]{Nienke van der Marel}{11},
  \au[0000-0002-5352-2924]{Sebastián Marino}{16},
  \au[0000-0002-7453-2945]{Maksym Mohorian}{1},
  \au[0000-0001-5907-5179]{Christophe Pinte}{19},
  \au[0000-0002-4716-4235]{Daniel Price}{20},
  \au[0000-0003-3818-408X]{Laurent Pueyo}{21},
  \au[0000-0003-2259-3911]{Shrishmoy Ray}{1},
  \au[0000-0002-9372-5056]{Nour Skaf}{22},
  \au[0000-0002-8942-1594]{Floris van der Tak}{23},
  \au[0000-0001-7026-6291]{Peter Tuthill}{8},
  \au[0000-0002-4147-3846]{Miguel Vioque}{17},
  \au[0000-0003-0774-6502]{Jason Wang}{5}
  \& \au[0000-0002-8537-9114]{Shangjia Zhang}{24}}

\coveraffiliations{%
  $^{1}$Macquarie University;
  $^{2}$California Institute of Technology;
  $^{3}$Center for Astrophysics \textbar{} Harvard \& Smithsonian;
  $^{4}$Stony Brook University;
  $^{5}$Northwestern University;
  $^{6}$Max-Planck-Institut f\"ur Astronomie;
  $^{7}$CNRS/IPAG, Grenoble;
  $^{8}$University of Sydney;
  $^{9}$University of Virginia;
  $^{10}$Commonwealth Scientific and Industrial Research Organisation;
  $^{11}$Leiden University;
  $^{12}$University of Victoria;
  $^{13}$Amherst College;
  $^{14}$University of Galway;
  $^{15}$University of Georgia;
  $^{16}$University of Exeter;
  $^{17}$European Southern Observatory;
  $^{18}$UMBC / NASA Goddard;
  $^{19}$Independent;
  $^{20}$Monash University;
  $^{21}$Space Telescope Science Institute;
  $^{22}$Institute for Astronomy, University of Hawaii;
  $^{23}$SRON Netherlands Institute for Space Research;
  $^{24}$Columbia University, NHFP Sagan Fellow}
  
\covercontact{%
  $^{*}$Joint first authors:
  \href{mailto:benjamin.pope@mq.edu.au}{benjamin.pope@mq.edu.au},
  \href{mailto:yinuo@caltech.edu}{yinuo@caltech.edu}}
\coverfigure[3.4in]{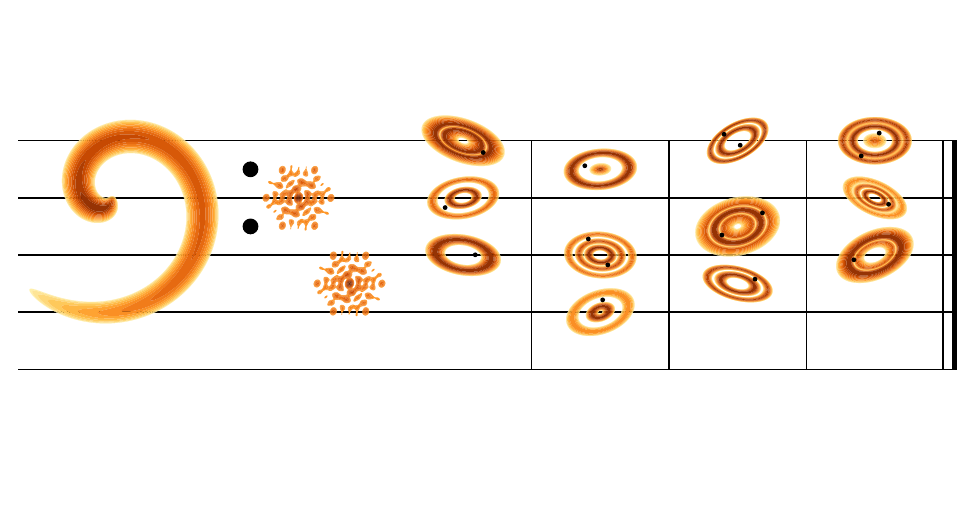}

\shorttitle{\textcolor{coverorange}{DMAJOR Legacy Survey of Protoplanetary Disks}}
\shortauthors{\textcolor{coverorange}{Pope, Han, et al.}}

\colorlet{COVERORANGE}{coverorange} 

\begin{document}

\title{\textcolor{coverorange}{DMAJOR: A JWST Legacy Survey of Protoplanetary Disks}}
\makecoverpage

\begin{abstract}
We propose an infrared imaging survey of resolved protoplanetary disks spanning a range of galactic environment, directly mapping the 3D structure of planet-forming disks across the IR. DMAJOR (Disk Motions and Architectures JWST Observational Reference survey), our legacy survey, will deliver a benchmark sample to reveal structure and connection to environment. The exquisite stability of JWST allows us to probe for variability with a time-scheduled revisit strategy, while its sensitivity and resolution open a window on disks unobservable from the ground. A uniform sample will reveal how small grains and ice are distributed in protoplanetary disks, how scattered light structures relate to large-scale environmental features, and identify new companions. To achieve this, DMAJOR takes advantage of model-based data analysis using telemetry and mutual calibration to enhance processed data products while minimizing telescope overheads.
\end{abstract}
\vspace*{-1.3cm}
\section{Scientific Justification}
\label{sec:justification}

Current understanding of the structure and morphological diversity of protoplanetary disks is limited by an observational bottleneck: there is no uniform population of resolved infrared images recovered at sufficient sensitivity. This survey fills this gap using the \ac{jwst} to produce a legacy image sample of resolved protoplanetary disks.
As the first large survey to cover all known, accessible, resolved protoplanetary disks with JWST/NIRCam coronagraphy and NIRISS/AMI interferometry, coverage extends further to the red than ground-based NIR imagers like VLT/SPHERE \citep{Beuzit2019} while delivering very high contrasts. This combination of filters and angular scales makes this regime the critical missing link to ALMA surveys like DSHARP and exoALMA \citep{Andrews2018,Teague2025}. The NIRCam coronagraph is the best instrument to spatially resolve faint scattered light from these disks at wavelengths longer than 2\,\micron, and yet its wide inner working angle means that fewer than 100 known, resolved Class~II disks are straightforwardly accessible to it across the sky. To comprehensively study these with NIRCam coronagraphy and complementary NIRISS/AMI imaging would cost around 700\,h of JWST time; the disks in the nearby Sco-Cen and Taurus associations around 300\,h. The Galaxy has provided us with a complete sample and a compelling science case naturally matched to the scale of a JWST Multi-Cycle Treasury Program!

\subsection{Open Questions about Protoplanetary Disks}
\label{sec:science}

Four decades \citep{Smith1984} of disk imaging campaigns have transformed our understanding of the origins of planetary systems; however, this knowledge is primarily informed by a small subset of bright disks that are at the focus of a sustained observational spotlight \citep{Hughes2018, Bae2023, Benisty2023, Ginski2024, Garufi2024}. With JWST/NIRCam's sensitivity and PSF stability offering unprecedented access to faint disk emission, we finally have the capability to deliver an expansive observational stage encompassing an all-sky survey of planet-forming disks.

\vspace{0.1cm}
\textit{(1) How are small grains distributed?} $\ $ While ALMA has shown that mm pebbles vertically settle to the disk midplane and are radially more compact than the gas disk \citep{Ansdell2018, Trapman2019, Oberg2021}, small grains are expected to follow the gas distribution, with any deviations suggesting local dust depletion or the motion of dust particles decoupled from gas \citep{Duchene2024, Ginski2024}. As the sensitivity limit of ground-based observations often truncates faint outer disks \citep{Rich2020}, IR observations sensitive to peripheral disk regions are required to robustly compare the small grain distribution with that of gas and pebbles \citep{Benisty2023, Duchene2024}.

\vspace{0.1cm}
\textit{(2) How is water ice distributed on the disk surface?} $\ $ Water ice is challenging to characterize with ground-based observations due to telluric lines, and it has only been mapped in a small number of protoplanetary disks \citep{Honda2016, Betti2022}. A wider resolved sample of surface water ice will assess the water reservoir for planets and the effects of stellar radiation and vertical mixing central to disk evolution \citep{Dartois2025, Ballering2025}. 

\vspace{0.1cm}
\textit{(3) How does the small grain and ice distribution relate to the large-scale environment and evolutionary stage?} $\ $ Extensive ALMA observations have shown significant evolution in the spatial distribution of pebbles from embedded disks \citep{Ohashi2023, Hsieh2025} to Class~II protoplanetary disks \citep{Andrews2018, Teague2025} to mature debris disks \citep{Matra2025, Marino2026}, while also revealing how environmental factors including late infall (streamers) and external photoevaporation shape fundamental disk properties \citep{Ginski2021, Winter2022, Aru2024}. The IR landscape remains statistically confined \citep{Garufi2026}. An IR survey will link the midplane evolution with that of the disk surface to construct a three-dimensional view, and JWST's sensitivity and large field of view will enable simultaneous characterization of large-scale structures such as streamers that feed the disk.

\vspace{0.1cm}
\textit{(4) Are disk structures shaped by planets?} $\ $ While dozens of companions have been directly imaged in the outer regions of mature planetary systems \citep{Vigan2021, Currie2023}, young planets captured during formation remain severely limited in number \citep[e.g., PDS~70 and WISPIT~2;][]{Keppler2018, vanCapelleveen2025}. An all-sky disk survey will search for wide-orbit companions and connect their presence (or absence) to the disk structure, evolutionary state, and formation process \citep{Bae2023}. Our sample will subsequently be targeted with VLTI and Keck/SCALES to reveal inner disks around the primary, exploring the physics of (mis-)alignment between the inner and outer disk and tracking accretion tracers such as Br$\gamma$ and Br$\alpha$.

\vspace{0.1cm}
\textit{(5) How common are IR substructures and how do they vary over time?} $\ $ Ground-based NIR observations have revealed that a subset of the brightest protoplanetary disks exhibit time-variable illumination linked to shadows from a dynamically active inner disk, with timescales of variability spanning from days \citep{Pinilla2018} to years \citep{Debes2017}. While this offers an exciting opportunity to probe inner disk dynamics otherwise unresolvable, a robust monitoring campaign requires non-uniform cadence and very high instrumental stability to confidently characterize well-detected shadowed disks identified in the initial survey. 

\subsection{State of the Art in High-Contrast Imaging}
\label{sec:pipelines}

In the workhorse data analysis pipelines for high contrast imaging, differential imaging methods are generally used to build a model of the stellar \ac{psf} from ensembles of images, which then are subtracted from the original images to find planets in residual maps \citep{follette2023,Cantalloube2021,Kenworthy2025}. This often involves \ac{rdi} with respect to a known point source, or \ac{adi} observing the same star at different roll angles, and taking optimized linear combinations of these to achieve maximal speckle subtraction while minimizing attenuation of astrophysical image components \citep[e.g.,][]{Lafreniere2007,Soummer2012,Pueyo2016,Ren2018}.
These approaches have been taken in \ac{jwst}/NIRCam imaging surveys to recover images and sensitive upper limits for disks \citep[e.g.,][]{Mullin2024,Gaspar2026} and point sources \citep[e.g.,][]{Bogat2025, Balmer2025b,Balmer2026}, and have permitted discovery and characterization of planets at unprecedented longer wavelengths with MIRI \citep[e.g.,][]{Malin2024,Malin2025}.

Recent approaches to space-telescope data analysis have taken a different approach: instead of reducing data and producing difference images, a physical forward model simulation is fit to the data end-to-end with free parameters for the calibration state and the astrophysical scene. This has been applied successfully to the NIRSpec IFU \citep{Ruffio2024} and JWST/NIRCam \citep{Balmer2025} using \ac{psf} models from \stpsf{} \citep{webbpsf_2012,webbpsf_2014}.

Profound gains have recently been demonstrated from models constructed in a differentiable framework such as \jax{} \citep{jax} or PyTorch \citep{pytorch}, which make it possible to optimize or sample models in arbitrarily high dimensions \citep{backprop,deep_learning}. For example the NIRISS \ac{ami} masking interferometer, whose commissioned sensitivity was more than an order of magnitude away from the projected photon-noise limit \citep{sallum_2024,ray2025} (a problem primarily due to the Brighter-Fatter Effect \citep[BFE,][]{bfe_ccd,bfe_ccd2,bfe_miri,bfe_fgs,bfe_pipeline}) has recently been recalibrated employing differentiable codes. This led to the development of \amigo{} \citep{Desdoigts2025,charles2025}, a forward model of \ac{ami} combining a differentiable optical simulation \citep{dlux1, dlux2} with a neural network \ac{edm} of the \ac{bfe}. By including the joint effects of these optical and electronic systematics, it has now been shown to reach close to the diffraction limit in resolution and to the photon noise limit in sensitivity. Similar approaches have been applied to NIRCam wavefront sensing \citep{Ray2026}, and extended to coronagraphy on JWST \citep{Feng2025} and HST/STIS \citep{Cao2026}.

JWST high contrast imaging techniques can therefore now achieve high image quality very close to the inner working angle, escaping limitations of the BFE and PSF miscalibration, and there is an emerging ecosystem of tools and approaches making use of these modern ideas. Crucially, forward-model calibrations can be learned hierarchically from an ensemble of sources without need for observation of dedicated \ac{psf} reference stars or injected diversity (such as various roll angles), and can flexibly incorporate information from wavefront metadata \citep{DeFurio2025}.

\subsection{Survey Strategy}
\label{sec:strategy}

\textbf{Sample}: We wish to conduct a survey of all the protoplanetary disks accessible to NIRCam and NIRISS/AMI: this will include all Class~II protoplanetary disks, excluding edge-on systems (which do not require coronagraphy), with estimated outer radii $\gtrsim 1''$ (1.5 times the inner working angle of the NIRCam MASK335R coronagraph). This mask is chosen to balance inner working angle with performance in the long-wavelength channel where JWST's reach is novel. Radii are inferred from $^{12}$CO \citep[from][]{Ansdell2018, Long2022, Sanchis2021, Stapper2024, Longarini2025, Deng2025, RuizRodriguez2025}, mm continuum surveys \citep[from][]{Andrews2018, Huang2018, Hendler2020, Francis2020, Cieza2021, Stapper2022}, or scattered-light surveys with SPHERE or GPI \citep{Monnier2017, Monnier2019, Avenhaus2018, Garufi2018, Garufi2020, Garufi2024, Laws2020, Rich2022, Ginski2024, Valegard2024, Byrne2026}, with a special focus on targets known to be members of the Taurus \citep[as documented in][]{Luhman2023} or Upper Sco/Sco-Cen associations \citep[from][]{Luhman2022, Luhman2025}. We exclude disks already studied extensively with 3--5\,\micron\ JWST coronagraphy \citep[e.g.,][]{Wagner2024, Uyama2025, Lin2026}, but include ten sources where one component of the DMAJOR survey strategy has been observed but others are missing, in order to obtain a complete and uniform sample. Based on these cuts, we find that there are 24 disks in Taurus, 20 in Sco-Cen (16 in Upper Sco and 4 in Upper Centaurus--Lupus and Lower Centaurus--Crux), and 47 other targets (16 in Lupus, 13 in Ophiuchus, 9 in Chamaeleon, 2 in Orion, 2 in Corona Australis, and 5 field stars), for a total of 91, of which 24 have a documented stellar companion within 10''. In finalizing proposed target lists towards survey execution, these numbers will be updated to reflect the most recent state of knowledge.

\textbf{Benchmark Systems}: This sample includes well-characterized disks with exoALMA (e.g., LkCa~15, HD~135344B, SY~Cha) \citep{Teague2025, Longarini2025} and MAPS (e.g., GM~Aur, MWC~480) \citep{Oberg2021} for robust comparison with ALMA observations. This sample also includes a subset of systems that have been studied by at least one other frontier instrument such as VLT/SPHERE or Gemini/GPI \citep{Avenhaus2018, Garufi2024, Ginski2024, Valegard2024, Rich2022} in addition to ALMA, but not by \ac{jwst}; or which have been partially studied with \ac{jwst} and merit multiwavelength follow-up.

\textbf{Instruments}:  To achieve the desired contrasts with NIRCam, observations will employ coronagraphy with a single roll angle using the MASK335R occulter. With its $\sim0.6''$ inner working angle, a considerable portion of the inner system is obscured behind the occulting spot. This will be filled in with complementary AMI data which can resolve sources down to $0.1''$ and has an outer working angle approaching $1''$. 
For NIRCam, we use 3 sets of short and long wavelength filter combinations (F444W+F200W, F300M+F210M and F360M+F182M), covering the 3\,\micron\ water ice feature with F300M and continuum filters on either side (F210M and F360M). 
All embedded companions discovered will motivate follow-up with the NIRSpec IFU to measure spectra and test for common proper motion, and would make ideal targets for characterization with the ELT.

\textbf{Time Domain:} We will image all targets once initially. An initial analysis of this first survey epoch will triage targets worth time-domain follow-up or characterization with alternative instruments, focusing particularly on embedded planets and self-shadowing features. For targets with self-shadowed features, we will follow up with a mix of short (days) and long (months to years across multiple JWST cycles) cadences sensitive to a range of plausible dynamical timescales of the inner disk, drawn from a fixed budget of epochs with most of the observing time concentrated in the first year.

\textbf{Calibration}: With a program of this scope, dedicated \ac{psf} reference stars per target add an unnecessary overhead to time, which is better spent observing targets of interest. We will request observation epochs contemporaneous with the regular wavefront sensing and maintenance cycles to maximize calibration quality with our forward model-based pipelines. Target visits are batched into small regions of the sky so that they share an instrument state and can be pooled with a small number of \ac{psf} reference stars for each group. All coronagraphic or AMI sources will be observed with 5 or 25 subgrid dithers to maximize the model-dependent registration of the star under the mask. Use of a range of filters allows for use of spectral information in deconvolution. 

\textbf{Software and Data Management}: Before the beginning of the program, we distill data from all available \ac{psf} reference stars and wavefront sensing data into accurate, differentiable forward models of each instrument, simultaneously benchmarking the performance of our pipelines on published targets. This workflow is already in progress or complete for NIRISS/AMI \citep{Desdoigts2025}, NIRCam coronagraphy \citep{FerrerChavez2026}, and HST/STIS \citep{Cao2026}, and in future \textit{Roman} \citep{Wang2026}. We will publish all software as documented, unit-tested open-source pipelines on GitHub, and in publications with reproducible workflows. Final images, spectra, time-series, and intermediate calibrations will be provided (hosted on MAST as high-level science products).

\section{Estimated Observing Time Required}
\label{sec:observing_time}

In most disk imaging scenarios, the achievable contrast is set by residual starlight, background, and readout noise rather than stellar photon noise, and observing time requirements are similar per target across a wide range in distance and flux. By analyzing the full set of group/integration/exposure settings from archival data, we find that NIRCam coronagraphic observations cost around $2150^{+2460}_{-890}$\,s of exposure per filter per roll across 42 programs, or $0.70^{+0.8}_{-0.3}$\,h of charged time per filter per roll. 

For AMI imaging, we follow the prescription that achievable contrast scales as $c \sim 100/\sqrt{N_\text{photons}}$ \citep{Desdoigts2025}, choosing $10^9$ photon integrations in each of the three available filters so as to be sensitive to bright point-source companions at $\Delta\text{mag} \sim 8.75$. 

We will minimize the fixed overheads of slewing to the PSF reference stars by selecting targets grouped together on sky and pooling science targets to PSF references in a 2:1 ratio. Model-based calibration (rather than KLIP or ADI) obviates the need for multiple roll angles, dramatically reducing calibration overheads.

For targets in Taurus or Sco-Cen, this prescription requires $\approx 5.6$\,h per star (3.6\,h of single-roll, three-color NIRCam coronagraphy with model-based PSF subtraction, plus NIRISS AMI on the 12 hosts brighter than $F_{W1} = 0.2$\,Jy, with PSF reference stars pooled at one per two science targets and all overheads included). A full program limited to a nominal 300\,h delivers all of the Taurus/Upper Sco Class~II disks, with $\sim$100\,h to spare for follow-up of targets of interest.

Completion of first pass visits for the full sample requires $\approx 330$\,h in NIRCam (91 disks, three filter pairs, single roll) and $\approx 230$\,h in AMI observations (44 hosts), looking only at appropriately bright targets ($F_{W1} \geq 0.2$\,Jy). Assuming that 10\% of these require follow-up with NIRSpec for embedded companions ($\approx 75$\,h), and 10\% with further NIRCam coronagraphic epochs to study motions ($\approx 56$\,h for two further epochs), we estimate that $\approx 690$\,h are required to survey the entire sky's accessible population of resolved disks.

\newpage
\bibliographystyle{whitepaper}
\bibliography{ms}

\end{document}